# Magnetic and Moiré Proximity Effects in WSe$_2$/WSe$_2$/CrI$_3$ Trilayers

Junyi Liu[†], Xu Zhang[†], and Gang Lu[*,†]

[†]*Department of Physics and Astronomy, California State University Northridge,*

*California 91330-8268, United States*

Corresponding author: Gang Lu
**Email:** ganglu@csun.edu.

**ABSTRACT**

Integrating magnetic order to moiré superlattices is of significant scientific and technological interest. Based on first-principles calculations, we study the interplay of magnetic proximity and moiré proximity in WSe$_2$/WSe$_2$/CrI$_3$ trilayers with different stackings and twist angles. Large valley splitting is observed due to redistribution of the exciton charge density across layers via a super-exchange-like mechanism, and its electric-field dependence bears similarity to electrically tunable and valley-selective Feshbach resonances. The valley splitting can be magnified in moiré superlattices owing to the superposition of Umklapp excitons folded from moiré minibands, yielding spatially modulated and enhanced magnetic proximity. The moiré proximity effect is demonstrated via an imprinted moiré potential on CrI$_3$ layer and its feedback to the direct moiré potential on WSe$_2$ bilayers is observed. The cooperation between the direct and imprinted moiré potentials is shown to yield novel topological and correlated states.

## I. INTRODUCTION.

Magnetic proximity effect is essential to diverse physical phenomena and technologies, enabling integration of magnetic heterostructures with superconducting, spintronic, topological and excitonic materials [1-4]. Recent discovery of two-dimensional (2D) magnets [2,5] opens possibilities of incorporating magnetic order to van der Waals (vdW) materials. For example, extensive experimental and theoretical research has been conducted to explore the magnetic proximity effect in vdW heterostructures between monolayers of transition metal dichalcogenides (TMDs) and 2D ferromagnets, including manipulation of spin and valley degrees of freedom in TMD monolayers[6-10], spin-dependent charge transfer [11], optical control of magnetization [12], and exciton valley splitting and polarization in the heterostructures [4,13-15], without external magnetic fields.

Moiré superlattices, formed by stacking 2D crystals with a small twist angle or lattice mismatch, provide a powerful and versatile platform to discover and engineer quantum materials [16,17]. Semiconducting moiré superlattices formed by TMDs are of particular interest because they can host moiré excitons which are correlated electron-hole pairs trapped at high-symmetry moiré lattice sites with large binding energies [18-20]. The moiré excitons can undergo Bose-Einstein condensation [21-23] and are a subject of enticing applications, from single-photon emitters [24] to optoelectronic [25] and excitonic devices [26]. One exciting and potentially fruitful research direction is to integrate magnetic order to moiré superlattices via the magnetic proximity effect. However, very little research has hitherto been devoted to moiré superlattices as recent effort has mainly focused on magnetic heterostructures with TMD monolayers [27,28], precluding the formation of moiré superlattices.

Notably, the proximity effect has been applied to moiré superlattices, but in a different context. A so-called moiré proximity effect has been proposed recently to imprint moiré patterns onto a proximal target "moiré-free" material, realizing correlated phases and unconventional excitonic states in the target [29-32]. Success has been achieved on various targets, including monolayer $MoSe_2$ [30], $WSe_2$ [31], and bilayer graphene [32] with imprinted moiré potentials produced by twisted TMD and graphene bilayers. However, the moiré proximity effect has not been demonstrated on magnetic target layers. Moreover, as envisioned in a commentary paper [29], the most exciting possibility is that the correlated phase in the target also exerts feedback onto the moiré superlattice, yielding "unexpected inter-target-substrate cooperative behavior".

In this work, we explore the magnetic proximity and moiré proximity effects in a single system, i.e., twisted $WSe_2/WSe_2/CrI_3$ trilayers, based on large-scale first-principles calculations. The magnetic proximity is effected by a monolayer $CrI_3$ which is a ferromagnetic semiconductor with out-of-plane magnetic moments [5], while the moiré proximity effect is induced by a twisted $WSe_2$ bilayer which imprints its moiré potential onto the proximal $CrI_3$ layer. Thus, in a single system we can examine both the magnetic and moiré proximity effects by focusing on the interplay of photon, exciton and magnetism therein.

Although first-principles calculations could provide unique insights beyond the reach of experiments and phenomenological models, they can be too expensive to be carried out for a moiré superlattice with thousands of atoms in the unit cell. This is particularly true when many-body excitonic effects are considered along with noncollinear magnetism and spin-orbit coupling. In this work, we overcome the computational challenges by employing a recently developed linear-response time-dependent density functional theory (TDDFT) method [33] with optimally tuned, screened and range-separated hybrid functionals [34-37]. Based on spinor wavefunctions, the TDDFT method can treat noncollinear magnetism and spin-orbit coupling with thousands of atoms in a unit cell, and it has been used to study moiré magnetism and moiré excitons in twisted CrSBr [38], TMD [39-41] and phosphorene [42] bilayers.

     Using the first-principle method, we have examined the magnetic and moiré proximity effects in $WSe_2/WSe_2/CrI_3$ trilayers with different stackings and twist angles. The exciton valley splitting is shown to be stacking and electric-field dependent, analogous to electrically tunable, valley-selective Feshbach resonances discovered in $MoSe_2$ bilayers [43,44]. The large valley splitting is due to redistribution of the exciton charge density across layers via a super-exchange-like mechanism. The valley splitting could be magnified in moiré superlattices thanks to the superposition of Umklapp excitons folded from moiré minibands, yielding spatially modulated and enhanced magnetic proximity effect. We have calculated the imprinted moiré potential on $CrI_3$ layer and examined its effects on excitons. Remarkably, we observe the envisioned feedback from the target layer ($CrI_3$) onto the moiré superlattice (twisted $WSe_2/WSe_2$). The "unexpected inter-target-substrate cooperative behavior" is shown to displace the moiré excitons to a new set of positions forming a Kagome lattice, which could in turn lead to interesting topological and correlated states.

## II. RESULTS AND DISSCUSSION

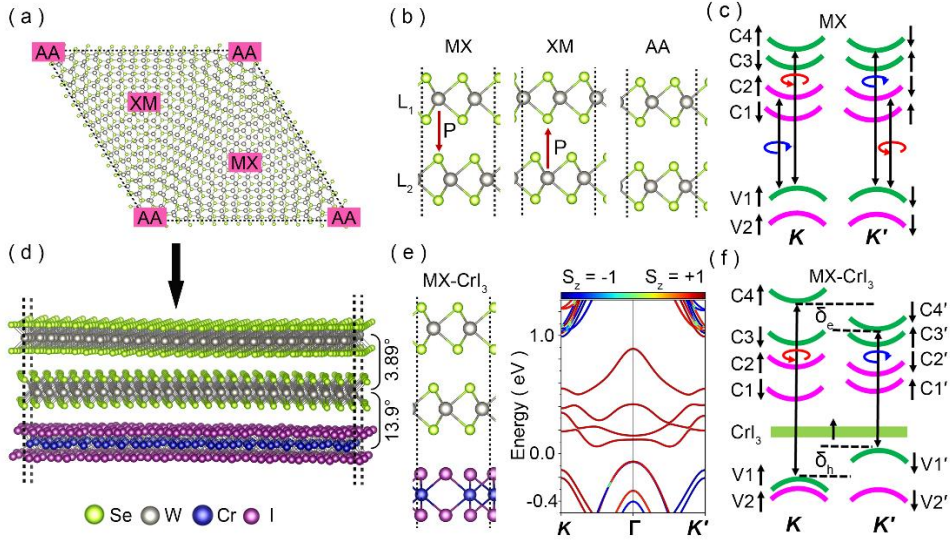

**Fig.1** (a) Top view of the twisted WSe₂ bilayer (θ=3.89°) with three high-symmetry stacking motifs highlighted. (b) Side view of the three stacking motifs with an electric dipole P. (c) Energy diagram of MX-stacked WSe₂ bilayer at $K$ and $K'$ valleys, where the energy levels from the bottom (L₂) and top (L₁) WSe₂ layer are shown in green and magenta color, respectively (the spin polarization is represented by a single-sided arrow). Electronic transitions (double-sided arrows) between the valence bands (v) and conduction bands (c) are coupled to the counterclockwise ($\sigma^+$, red) and clockwise ($\sigma^-$, blue) circularly polarized photons. (d) Side view of the twisted WSe₂ bilayer (θ=3.89°) on a CrI₃ monolayer with a relative twist angle of 13.9° between WSe₂ layer (L₂) and CrI₃. (e) Side view of the non-twisted trilayer MX-CrI₃ with MX-stacked WSe₂ bilayer on top of CrI₃ monolayer (left panel). The band structure of MX-CrI₃ with spin polarization indicated by the horizontal color bar. The Fermi level is set at zero energy (right panel). (f) Energy splitting at $K/K'$ valley in MX-CrI₃ with $\delta_e$ and $\delta_h$ denoting the electron and hole contribution to the valley splitting. The horizontal green bar represents CrI₃ layer with spin-up electrons.

In **Fig. 1(a)**, we show the top view of a twisted WSe₂ bilayer with a twist angle θ=3.89°. The three high-symmetry stacking motifs (MX, XM, and AA) are highlighted with their atomic structures illustrated in **Fig. 1(b)**. In the MX stacking, the metal atoms in L₁ layer sit on top of the chalcogen atoms in L₂ layer, yielding a dipole P with an electric field of -62 mV/nm. An opposite electric field (+62 mV/nm) is present in the XM stacking. AA stacking carries no vertical dipole or electric field. The electric field can influence the magnetic proximity effect as discussed below. We first establish spin-valley-layer locking and optical selection rules in non-twisted WSe₂ bilayers. Owing to spin-orbit coupling and time-reversal symmetry, $K$ and $K'$ valleys of WSe₂ bilayer have degenerate bands with opposite spin orientations, as shown in **Fig. 1(c)**. In each WSe₂ layer, the intralayer A exciton at ***K*** valley couples to σ⁺ photons via a single-particle transition V1→C4 whereas at ***K′*** valley it couples to σ⁻ photons. The interlayer excitons, on the other hand, may exhibit opposite selection rules to the A exciton [45]. For example, an interlayer exciton at ***K*** valley couples to σ⁻ (V1→C2) as opposed to σ⁺ for the A exciton. A full summary of the selection rules is given in **Table S1**.

We next examine the magnetic proximity effect on the non-twisted WSe₂ bilayers when a CrI₃ monolayer is placed adjacent to the WSe₂ bilayer as shown in **Fig. 1(e)**. The spin-resolved band structure of WSe₂/WSe₂/CrI₃ trilayer with MX stacking between WSe₂ layers is shown in **Fig. 1(e)**. The four lowest conduction bands are derived from CrI₃ and spin-polarized ($S_z=1$), whereas the top valence bands are from WSe₂ with opposite spins at ***K*** and ***K′*** valleys. More importantly, the energy degeneracy at the two valleys is broken due to the magnetic proximity effect, giving rise to valley Zeeman splitting as illustrated in **Fig. 1(f)**. In a typical experiment, the valley splitting Δ is measured as the energy difference between the optical absorption peaks from σ⁺ and σ⁻ polarized photons [4,15], *i.e.*, $\Delta = E(\sigma^+) - E(\sigma^-)$, where $E(\sigma^+)$ and $E(\sigma^-)$ are the absorption energies of σ⁺ and σ⁻ photons. The absorption energies can be calculated from first-principles by incorporating many-body electron-hole correlations to account for the excitonic effect, which is important in 2D materials. However, the many-body calculations (e.g., GW-BSE) are computationally intensive, thus in practice single-particle calculations are widely used to examine magnetic proximity in 2D vdW heterostructures [7-10,27,28]. In this work, we carry out both TDDFT and single-particle DFT+U calculations, with the former capable of capturing the excitonic effect. From the single-particle calculations, $E(\sigma^+)$ and $E(\sigma^-)$ can be estimated by energy differences between the conduction and valence bands at ***K*** and ***K′*** valleys. For example, the valley splitting $\delta$ in **Fig. 1(f)** can be expressed as:

$$\delta = \left(E_{C4}(\boldsymbol{K}) - E_{V1}(\boldsymbol{K})\right) - \left(E_{C4'}(\boldsymbol{K'}) - E_{V1'}(\boldsymbol{K'})\right) = \left(E_{C4}(\boldsymbol{K}) - E_{C4'}(\boldsymbol{K'})\right) - \left(E_{V1}(\boldsymbol{K}) - E_{V1'}(\boldsymbol{K'})\right)$$
$$= \delta_e - \delta_h.$$

Here $E_{C4}(\boldsymbol{K})$, $E_{V1}(\boldsymbol{K})$, $E_{C4'}(\boldsymbol{K'})$, $E_{V1'}(\boldsymbol{K'})$ denote the conduction and valence band energies at ***K*** and ***K′*** valleys for the A exciton. Hence, the total valley splitting $\delta$ contains opposing contributions from electron ($\delta_e$) and hole ($\delta_h$); one can similarly define the valley splitting for interlayer excitons. $E(\sigma^+)$, $E(\sigma^-)$ and Δ can also be obtained from TDDFT calculations of optical absorption spectra as discussed later. Thus, by comparing the valley splitting between TDDFT calculations (Δ) and single-particle calculations ($\delta$), one can assess the excitonic effect on valley splitting.

Valley splitting is usually analyzed in terms of Zeeman effect, i.e., the energy splitting is linearly dependent on the magnetic field. And extensive research - both theory and experiment - has been devoted to deriving this linear coefficient, i.e., the *g* factor in 2D materials[4,46-48]. However, in this work, we find that the dominant contribution to the valley splitting in $WSe_2/WSe_2/CrI_3$ trilayers is actually interlayer hybridization that depends on the stacking and the internal electric field. In particular, we show that the electric-field dependence bears similarity to electrically tunable Feshbach resonances recently discovered in both twisted and non-twisted $MoSe_2$ bilayers [43,44].

To examine the stacking dependence of valley splitting, we carry out the single-particle calculations to determine $\delta$, $\delta_e$ and $\delta_h$ as a function of the interlayer displacement vector ***d*** defined in **Fig. 2(a)** with a primitive unit cell of $CrI_3$ and a (2×2) supercell of MX-stacked $WSe_2$ bilayer. The total valley splitting $\delta$ along with $\delta_e$ and $\delta_h$ as a function of ***d*** are displayed in **Fig. 2**. Four distinct energy maxima are observed in $\delta - $ ***d*** map and they correspond well to the minima in $d_{NN}$ - ***d*** map in **Fig. 2(c)** where $d_{NN}$ represents the shortest Cr-W distance. This correlation is made more clear in **Fig. 2(d)**. The valley splitting contribution from the hole ($\delta_h$) tracks very well with $\delta$, i.e., the same ***d*** yields the maximal $\delta$ and minimal $\delta_h$ simultaneously in **Fig. 2(f)**, suggesting that W 3*d*-band plays a critical role in Cr-induced hole valley splitting [7]. In contrast, the electron contribution $\delta_e$ does not correlate well with $d_{NN}$ (or $\delta$), implying that Se atoms may be responsible for $\delta_e$. In a twisted $WSe_2/WSe_2/CrI_3$ trilayer, ***d*** varies locally in the moiré superlattice, thus yielding spatially modulated valley splitting.

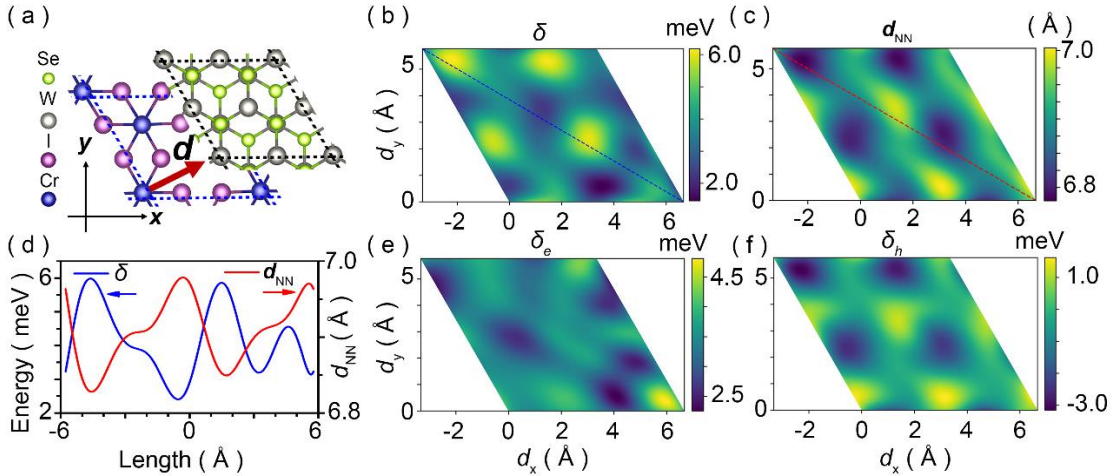

**Fig. 2** (a) The interlayer displacement ***d*** between the bottom ($L_2$) $WSe_2$ layer and $CrI_3$ layer defined in the unit cell of a non-twisted $WSe_2/WSe_2/CrI_3$ trilayer (the top $WSe_2$ layer is removed for clarify). (b) Total valley splitting $\delta$ as a function of $d$ ($d_x$ and $d_y$ are x and y components of $d$) determined from the single-particle calculations. (c) The shortest Cr-W distance ($d_{NN}$) as a function of $d$. (d) Total valley splitting $\delta$ and $d_{NN}$ along the dashed line in (b) and (c) with the midpoint set as zero. (e) Electron contribution to the valley splitting ($\delta_e$) as a function of $d$. (f) Hole contribution to the valley splitting ($\delta_h$) as a function of $d$.

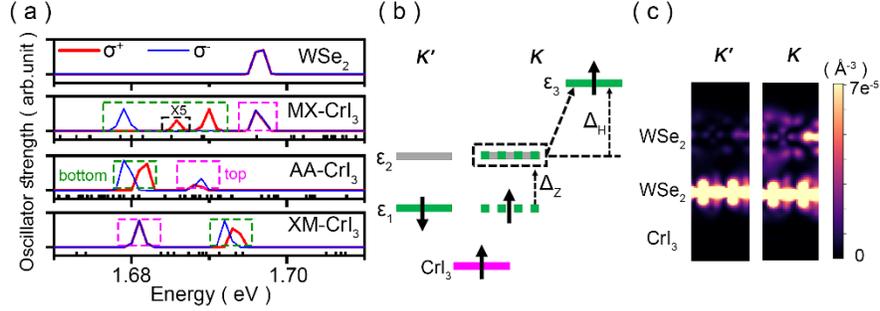

**Fig. 3** (a) Optical absorption for $\sigma^+$ (red) and $\sigma^-$ (blue) circularly polarized photons in a non-twisted $WSe_2/WSe_2/CrI_3$ trilayer with different $WSe_2$ stackings; $WSe_2$ monolayer is also included in the first row for comparison. The magenta/green box shows the absorption from the top/bottom $WSe_2$ layer. The satellite peak enclosed by the black dashed box in $MX$-$CrI_3$ is magnified by 5 times. (b) Schematic illustration of valley splitting contributions ($\Delta_Z$ and $\Delta_H$) at $K/K'$ valley in $MX$-$CrI_3$ trilayer. $\varepsilon_1$ represents the conduction electron states of the A exciton at the bottom $WSe_2$ layer whereas $\varepsilon_2$ represents the conduction state from the top $WSe_2$ layer at $K/K'$ valley. After hybridization with $\varepsilon_2$, the electron energy of the A exciton is increased to $\varepsilon_3$ at $K$ valley. (c) Side view of the electron density of the A exciton at $K'$ (left) and $K$ (right) valley for $MX$-$CrI_3$ trilayer.

We next perform TDDFT calculations to determine the optical absorption spectra for $WSe_2/WSe_2/CrI_3$ trilayers, examining how interlayer hybridization, internal electric field and excitonic effect may contribute to the valley splitting. To this end, we compare three non-twisted trilayers with different stacking configurations: $MX$-$CrI_3$, $AA$-$CrI_3$, and $XM$-$CrI_3$. In **Fig. 3(a)**, we display the absorption spectra of the three trilayers along with $WSe_2$ monolayer at an energy range close to the A exciton (~1.7 eV). The spectra are color-coded with the red (blue) curve corresponding to $\sigma^+$ ($\sigma^-$) photons absorbed at $K$ ($K'$) valley. A single prominent peak for the A exciton is featured in the absorption of $WSe_2$ monolayer, and this A exciton peak splits in the trilayers due to the presence of the internal electric field and the exchange field. In $MX$-$CrI_3$ and $XM$-$CrI_3$ trilayers, the internal electric field yields layer-selective splitting for the A exciton [49].

Specifically, the exciton energy at the bottom layer (green box) is lower than the top layer (magenta box) in MX-CrI$_3$ and vice versa in XM-CrI$_3$ trilayer. Although there is no internal electric field in AA-CrI$_3$ trilayer, the symmetry between WSe$_2$ layers is nonetheless broken because the bottom WSe$_2$ layer experiences a much stronger exchange field than the top layer (the exchange field decays rapidly with the distance). For this reason, in the following we will focus on the bottom WSe$_2$ layer to elucidate the magnetic proximity effect.

**Table 1.** Valley splitting for intralayer and interlayer excitons in the non-twisted WSe$_2$/WSe$_2$/CrI$_3$ trilayers. The contributions to valley splitting Δ from the atomic Zeeman ($\Delta_Z$), electric field ($\Delta_E$) and interlayer hybridization ($\Delta_H$) are listed separately (no corresponding data available for the interlayer excitons). The valley splitting ($\delta$) from the single-particle calculations and its electron ($\delta_e$) and hole ($\delta_h$) contributions are also included. The excitonic contribution is obtained as the difference between Δ and $\delta$.

|  |  | Δ | $\Delta_Z$ | $\Delta_E$ | $\Delta_H$ | $\delta$ | $\delta_e$ | $\delta_h$ | Δ-$\delta$ |
|---|---|---|---|---|---|---|---|---|---|
| Intralayer exciton | MX-CrI$_3$ | 11.1 | 2.2 | 1.2 | 7.7 | 5.4 | 7.3 | 1.9 | 5.7 |
| | AA-CrI$_3$ | 2.2 | 2.2 | 0 | 0 | 1.9 | 3.3 | 1.5 | 0.3 |
| | XM-CrI$_3$ | 1.0 | 2.2 | -1.2 | 0 | 0.9 | 3.0 | 2.0 | 0.1 |
| | | | | | | | | | |
| Interlayer exciton | MX-CrI$_3$ | -13.7 | - | - | - | -6.2 | -6.2 | 0 | -7.5 |
| | XM-CrI$_3$ | -4.8 | - | - | - | -2.6 | -2.6 | 0 | -2.2 |

In **Table 1,** we list the energy of the intralayer A exciton of the bottom WSe$_2$ layer for the three stackings, and we divide its valley splitting (Δ) into atomic Zeeman ($\Delta_Z$), internal electric field ($\Delta_E$) and hybridization ($\Delta_H$) contributions. The valley splitting ($\delta, \delta_e, \delta_h$) from the single-particle calculations is also included. Among the three trilayers, MX-CrI$_3$ has the largest valley splitting (11.1 meV) while XM-CrI$_3$ has the smallest (1.3 meV). This trend is supported by the previous finding that both valley splitting and quasiparticle bandgap of TMD/CrI$_3$ heterostructures scale approximately with the electric field [8], which in turn can be rationalized from second-order perturbation theory [27].

By assuming that all three trilayers have the same atomic Zeeman contribution ($\Delta_Z$) due to their identical atomic compositions, and that AA-CrI$_3$ trilayer has no electric field contribution ($\Delta_E = 0$) to the valley splitting, we can obtain $\Delta_Z$ and $\Delta_E$ for the three trilayers (**Table I**). It is revealed that the dominant contribution to the valley splitting in MX-CrI$_3$ is interlayer hybridization ($\Delta_H$ = 7.7 meV), which is surprising. To understand its origin, we present a schematic diagram in **Fig. 3(b)** showing the energy of the electron in the A exciton at *K/K′* valley of the bottom WSe$_2$ layer (the valley splitting due to the hole is less important, thus not considered). As mentioned earlier in **Fig. 1(e)**, the lowest conduction bands in MX-CrI$_3$ trilayer stem from CrI$_3$ and are spin-polarized as represented by the pink level in **Fig. 3(b).** The magnetic exchange interaction enables the mixture of these CrI$_3$ states with the spin-up electron in the A exciton at *K* valley of the bottom WSe$_2$ layer, raising its energy from $\varepsilon_1$ to $\varepsilon_1+\Delta_Z$. In contrast, there is little mixture between the CrI$_3$ states and the spin-down electron at *K′* valley. Note that this finding is consistent with spin-dependent interfacial charge transfer observed in magnetic heterostructures [11,14]. The spin-up electron is aligned energetically with $\varepsilon_2$ which originates from an unoccupied electron state at the top WSe$_2$ layer. The resonance of the two states leads to avoided crossing which further elevates the energy of the spin-up electron, yielding $\Delta_H$ contribution to the valley splitting. Notably, this avoided crossing is accompanied by electron tunneling from the bottom WSe$_2$ to the top WSe$_2$ layer, as shown in the electron density plot of the A exciton in **Fig. 3(c)**, with the appearance of interlayer electron states at *K* valley and the absence of such interlayer states at *K′* valley. In other words, the electron in the A exciton is partially transferred to the top layer and binds with the hole at the bottom layer to form an interlayer exciton, which presents itself as a satellite peak in the absorption of MX-CrI$_3$ at *K* valley ($\sigma^+$). The interlayer electron redistribution (with resultant electron-hole separation) reduces the exciton binding energy - making it less negative - and increases the exciton energy at *K* valley. In contrast, there is no such charge redistribution at *K′* valley. Thus, the exchange interaction with CrI$_3$ results in interlayer charge transfer of spin-up electrons at *K* valley. And a super-exchange-like interaction between the spin-up electrons in CrI$_3$ and the top WSe$_2$ layer, mediated by the electrons in the bottom WSe$_2$ layer, is responsible for $\Delta_H$ in MX-CrI$_3$. In contrast, there is no such interaction in AA-CrI$_3$ and XM-CrI$_3$ trilayers because their energy separations between ε$_1$ and ε$_2$ are too large for interlayer tunneling at ***K*** valley, hence $\Delta_\text{H}$ vanishes in these trilayers. In other words, the internal electric field in MX-CrI$_3$ enables electron tunneling across WSe$_2$ layers, producing long-range exchange and large valley splitting.

We can estimate the excitonic contribution to valley splitting, i.e., the difference between $\Delta$ and $\delta$ (Table 1). It is found that in MX-CrI$_3$ the excitonic contribution (5.7 meV) accounts for most of $\Delta_\text{H}$, confirming that the electron redistribution in the A exciton is mainly responsible for the large valley splitting. Moreover, the excitonic contribution is the greatest in MX-CrI$_3$ and negligible in AA-CrI$_3$ and XM-CrI$_3$ trilayers, suggesting that the super-exchange-like interaction in MX-CrI$_3$ is mediated by excitons, in line with the experimental observation that excitons could mediate long-range ferromagnetic exchange interactions between moiré-trapped holes in WS$_2$/WSe$_2$ bilayer [50].

Recent experimental and theoretical work has discovered electrically tunable Feshbach resonances in MoSe$_2$ bilayers [43,44]. It is found that in the presence of interlayer electron (or hole) tunneling, scattering of intralayer excitons and electrons (or holes) occupying different layers can be resonantly enhanced by tuning an applied electric field. More importantly, the Feshbach resonance is valley-selective and can induce exciton-mediated ferromagnetic interactions between the holes in moiré layers [51]. Remarkably, although in a very different context, our calculations also show hybridization between the intralayer A exciton and tunneling electron across the layers. This hybridization is also valley-selective (only at ***K*** valley) and depends on the electric field. In addition, similar avoided crossings are observed in the experiment and our calculations. The analogies imply that exciton valley splitting in the trilayer may share a similar physical origin as the Feshbach resonance.

There is also valley splitting in interlayer excitons. Here, we focus on an interlayer exciton in MX-CrI$_3$ and XM-CrI$_3$ whose energy is similar to the A exciton. The interlayer exciton has its electron/hole localized at the bottom/top WSe$_2$ layer (**Fig. S1**). As the hole is farther away from CrI$_3$ layer, its contribution to the valley splitting $\delta_h$ vanishes in MX-CrI$_3$ and XM-CrI$_3$ (**Table 1**). Hence, the total valley splitting $\delta = \delta_\text{e} - \delta_\text{h}$ of the interlayer exciton is larger than that of the intralayer exciton. The same reasoning also applies to $\Delta$ from the TDDFT calculations.

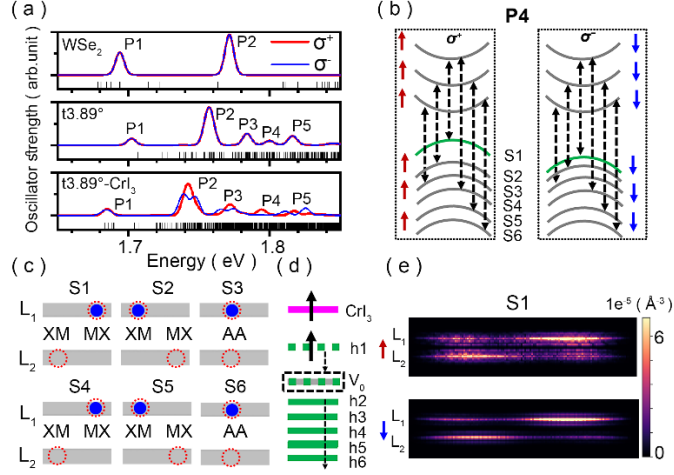

**Fig. 4** (a) Optical absorption for σ⁺ (red) and σ⁻ (blue) circularly polarized photons in WSe$_2$ monolayer (top), the twisted WSe$_2$ bilayer (θ=3.89°, middle) and the twisted WSe$_2$/WSe$_2$/CrI$_3$ trilayer (θ=3.89°, bottom). (b) The spin-up (red) and spin-down (blue) channels of single-particle excitations (S1 to S6) at $\gamma$ point of the moiré Brillouin zone for the twisted WSe$_2$/WSe$_2$/CrI$_3$ trilayer (θ=3.89°). (c) Schematic picture showing the position of the electron (filled circle) and the hole (open circle) for excitons S1 to S6. The high-symmetry points in each WSe$_2$ layer are indicated. (d) Schematic energy diagram for the spin-up holes at *K* valley in the twisted WSe$_2$/WSe$_2$/CrI$_3$ trilayer. Each green line represents the hole (h1 to h6) of the exciton (S1 to S6) which constitutes the P4 moiré exciton. The grey line represents the interlayer state V$_0$ of WSe$_2$ bilayer. The dashed arrow indicates $\Delta_Z$ contribution for each hole level. (e) The side view of the spin-up (top) and spin-down (bottom) hole density in S1.

We now turn to the magnetic proximity effect on moiré superlattices and elucidate the so-called moiré Zeeman effect. In **Fig. 1(d)**, we display the atomic structure of a twisted WSe$_2$/WSe$_2$/CrI$_3$ trilayer consisted of the twisted WSe$_2$ bilayer (θ=3.89°) on top of a CrI$_3$ monolayer, which is itself twisted relative to the bottom WSe$_2$ layer by 13.9°. The moiré unit cell of the trilayer has 1718 atoms with a lateral lattice constant of 4.9 nm. The valley splitting of the moiré trilayer can be obtained from the optical absorption spectra shown in **Fig. 4(a)**, including WSe$_2$ monolayer, the twisted WSe$_2$ bilayer (θ=3.89°) and the twisted trilayer. In WSe$_2$ monolayer, the P1 and P2 peaks correspond to the 1s and 2s states of the A exciton, respectively [6]; the 2s state has a larger Bohr radius and is more extended. In the twisted WSe$_2$ bilayer, P1 (P2) has four degenerate 1s (2s) A excitons, two at each WSe$_2$ layer. In addition, three bright exciton peaks (P3-P5) emerge at higher energies, and they are called Umklapp excitons because they originate from moiré minibands folded from the monolayer states near ***K*** and ***K′*** valleys via the Umklapp process [52].

In the twisted WSe$_2$/WSe$_2$/CrI$_3$ trilayer, the energy degeneracy of P1 and P2 peaks is broken by CrI$_3$, splitting both σ$^+$ and σ$^-$ peaks. The splitting of P2 is larger than P1 because the 2s wavefunction of P2 is more extended with larger overlap with CrI$_3$ wavefunction. However, the splitting of P2 σ$^+$ peak is smeared by its larger Zeeman splitting $\Delta_Z$ as shown in **Fig. 4(a)**. Thus, only P2 σ$^-$ peak shows clear splitting. The valley splitting of P1 and P2 is relatively smaller because as intralayer excitons, their electron and hole contributions to the valley splitting are partially cancelled. By contrast, the interlayer Umklapp excitons (P3-P5) exhibit much larger Zeeman splitting, with P4 having the largest splitting of -14 meV.

The moiré exciton at P4 is a superposition of six direct transitions (S1 to S6) at the *γ* point of the moiré Brillouin zone. Each *γ* exciton is represented by a dashed arrow in **Fig. 4(b)** with a spin-up and spin-down channel that couples to σ$^+$ and σ$^-$ photon, respectively. The overall valley splitting is a weighted sum from each *γ* exciton. These *γ* excitons are spatially localized as shown in **Fig. 4(c)** with distinctive valley splitting. For example, S1 exciton has the largest valley splitting of -51 meV while S2 and S3 have much smaller valley splitting of -0.7 and -0.6 meV, respectively. Hence, spatially modulated and enhanced magnetic proximity effect can be realized in the moiré trilayers, which could lead to programmable devices with spin functionality.

It turns out that the holes in these *γ* excitons play an important role in valley splitting, and in **Fig. 4(d)**, we present a schematic energy diagram for the spin-up holes in the *γ* excitons. The holes have negligible interlayer coupling since they originate from ***K/K′*** valleys of the monolayer [53]. Crucially, we identify an additional valence state V$_0$ in the energy range of the holes. Stemming from the *Γ* valley of the monolayer, V$_0$ has strong interlayer coupling and will be referred as the interlayer state in the following. Thanks to the exchange interaction with CrI$_3$, the energy of each spin-up hole is lowered relative to its spin-down counterpart, leading to $\Delta_Z$ contribution. However, as V$_0$ lies between h1 and the other holes, only h1 would move closer in energy to V$_0$ and hybridize with it. The hybridization – mixture of h1 and V$_0$ - further lowers the energy of h1 and gives rise to $\Delta_H$ contribution for S1. The other holes, on contrast, cannot hybridize with V$_0$, thus have much smaller valley splitting. Note that the hybridization can only occur in the spin-up channel, acquiring interlayer states. As shown in **Fig. 4(e)**, there is significant interlayer hole density in the spin-up channel of S1, but negligible hole density in the spin-down channel. The spin-dependent hole density redistribution is the origin of the large valley splitting in S1 (and P4). We thus conclude that the exciton charge density redistribution is chiefly responsible for valley splitting in the magnetic heterostructures – both twisted and non-twisted trilayers. Because a mixture of states tends to yield greater charge redistribution than tunneling, valley splitting in a moiré trilayer is usually larger than a non-moiré trilayer. In addition, a moiré exciton is a superposition of multiple excitons folded from moiré minibands. As a result, there are more states (e.g., h1 to h6) available in the moiré exciton to mix with the interlayer state for valley splitting. In contrast, there is only one state ($\varepsilon_1$) in MX-CrI$_3$ involved in tunneling. Therefore, moiré superlattices can magnify valley splitting. For example, giant valley splitting (~25 meV) was observed in WSe$_2$/WS$_2$ moiré bilayers [51]. In fact, bright Umklapp excitons in moiré superlattices have been proposed as direct optical probes for magnetic order in twisted TMD layers due to their enhanced Zeeman energy splitting [51,54].

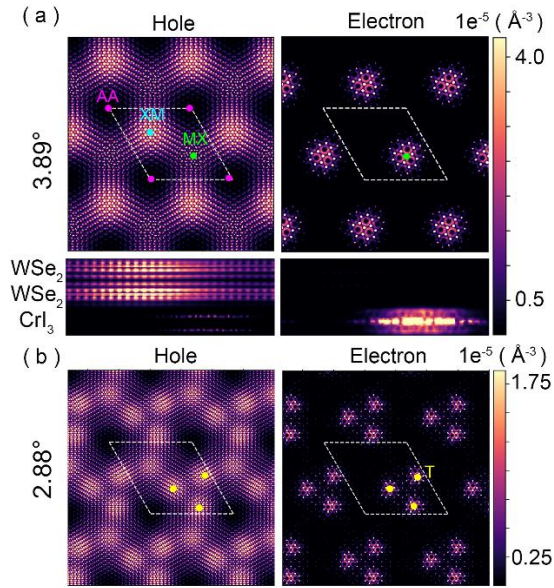

**Fig. 5** (a) Top and side view of the hole (left) and electron (right) density of the lowest-energy interlayer exciton in the twisted trilayer ($\theta$=3.89°). (b) Top and side view of the hole (left) and electron (right) density of the lowest-energy interlayer exciton in the twisted trilayer ($\theta$=2.88°). The position of "T" site is indicated by a yellow dot. In all figures, the charge density is color-coded, and the moiré unit cell is indicated by a dashed box with high-symmetry points labeled.

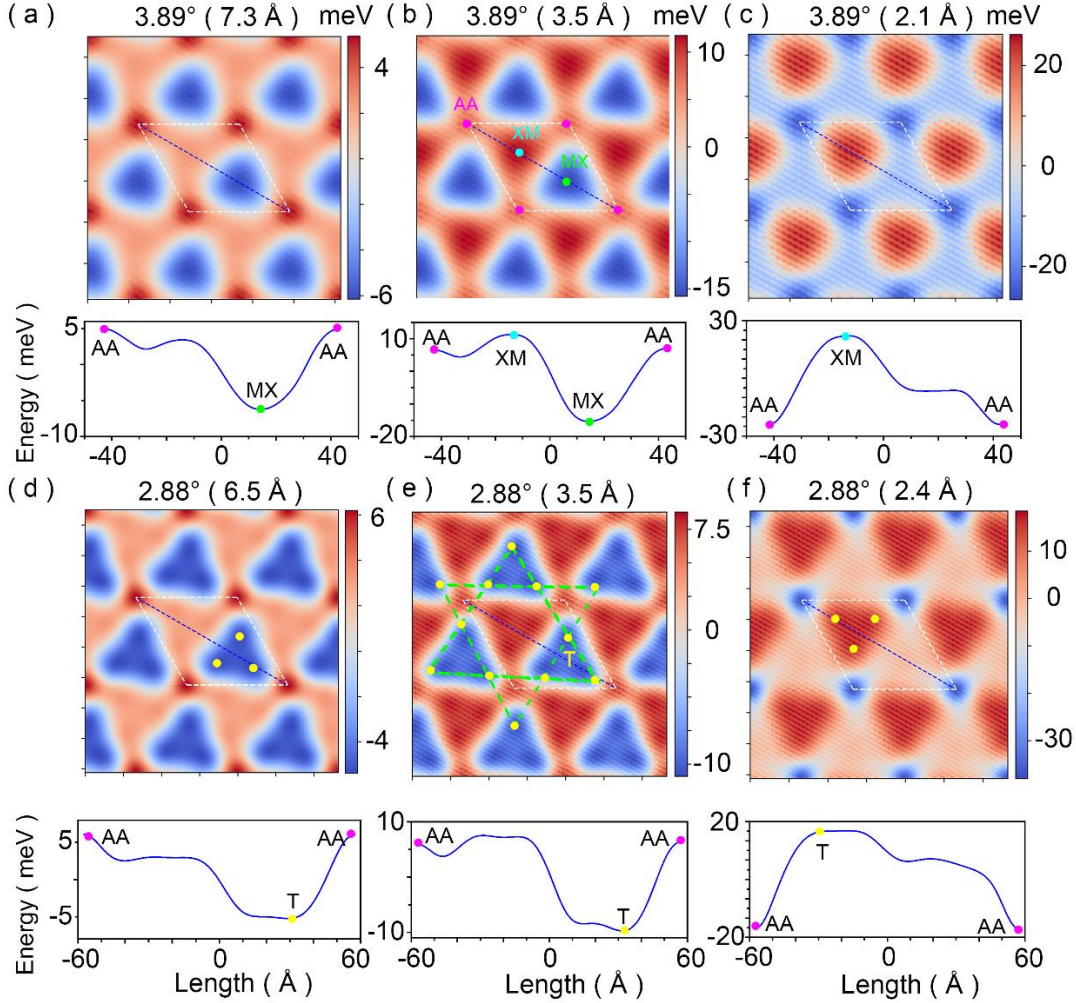

**Fig. 6** The imprinted moiré potential and its projection along the diagonal line of the moiré unit cell in a plane at 7.3 Å (a), 3.5 Å (b) and 2.1 Å (c) away from the twisted WSe$_2$ bilayer (θ=3.89°) and at 6.5 Å (d), 3.5 Å (e) and 2.4 Å (f) away from the twisted WSe$_2$ bilayer (θ=2.88°). The Kagome lattice formed by the "T" points (indicated by yellow dots) is shown in (e). Moiré unit cell is indicated by the white dashed box with the high-symmetry points labeled in (b).

Having studied how CrI$_3$ monolayer could affect moiré excitons in WSe$_2$ bilayers (i.e., magnetic proximity effect), we now explore how WSe$_2$ bilayers may conversely affect moiré excitons in CrI$_3$ monolayer (i.e., moiré proximity effect). More specifically, we examine how interlayer and intralayer excitons in CrI$_3$ may be influenced by the imprinted moiré potential yielded by the twisted WSe$_2$ bilayers.

In **Fig. 5(a)**, we show the charge density of the lowest interlayer exciton in $WSe_2/WSe_2/CrI_3$ trilayer ($\theta = 3.89°$). The hole resides at the MX and XM regions of the moiré $WSe_2$ bilayer and the electron at the MX region of $CrI_3$ layer. While the hole is confined by the direct moiré potential, the electron is trapped by the imprinted moiré potential. Interestingly, when the twist angle between $WSe_2$ layers changes to $\theta = 2.88°$ (**Fig. S2**), the trapped electron is displaced to three new positions labeled by "T" in $CrI_3$, deviating from the high-symmetry MX points. Remarkably, the hole responds by also shifting away from the high-symmetry MX points and moves toward the "T" points as shown in **Fig. 5(b)**, due to its Columbic binding to the electron. In other words, the direct moiré potential in $WSe_2$ bilayer is modified effectively, and there is feedback from the target layer ($CrI_3$) onto the moiré superlattice (twisted $WSe_2/WSe_2$) as envisioned in [29]. The feedback is expected to be stronger for a smaller $\theta$. The fact that the moiré exciton (both the electron and hole) is now localized at "T" points as opposed to MX points of the superlattice is a manifestation of the "unexpected inter-target-substrate cooperative behavior", which could have important consequences in engineering moiré materials. Since the imprinted potential depends on the twist angle and can be switched on/off by adjusting the vertical distance between the target and the source layers, one could thus tune the cooperative behavior on demand.

To be concrete, we calculate the imprinted moiré potential on a target plane, placed at a distance $h$ from the adjacent $WSe_2$ layer (the equilibrium distance $h_0 = 3.5$ Å for $\theta = 3.89°$). The imprinted potential is defined as the electrostatic potential (the sum of Hartree and local pseudopotential) on the target plane, in line with the experiments [29-32]. In **Fig. 6,** we show the imprinted potential (top) and its projection along the diagonal line of the moiré unit cell (bottom) for the two twist angles and three different $h$ values. For $\theta = 3.89°$ and $h = 3.5$ Å, we find that the imprinted potential reaches its minimum (-15 meV) at the MX point and its maximum (10 meV) at the XM point, as shown in **Fig. 6(b)**. This agrees with the observation that the electron of the interlayer exciton is trapped at the MX region in $CrI_3$ layer. For $\theta = 2.88°$ and $h = 3.5$ Å, the minimum of the imprinted potential shifts to the "T" points as shown in **Fig. 6(e)**, consistent with the result that the electron of the interlayer exciton is trapped at the "T" points in $CrI_3$ layer. Notably, the amplitude of the imprinted potential decreases from 25 meV ($\theta = 3.89°$) to 15 meV ($\theta = 2.88°$), following the opposite angle-dependence with the direct moiré potential. This highlights the distinction between the direct and imprinted moiré potentials: the former represents the energy variation of an exciton (charge neutral) while the latter of a charged carrier (electron or hole). **Fig. 6** also depicts the dependence of the imprinted potential on $h$. For example, the minimum of the potential could shift from the MX to AA points when $h$ is small (e.g., under a pressure). Conversely, when $h$ is large, the maximum of the potential shifts from the MX to AA points with a smaller potential amplitude (~10 meV). Importantly, we note that the energy minimal T points in θ =2.88° moiré trilayer ($h$ = 3.5 Å) form a Kagome lattice, implying a possible formation of topological flat bands for trapped electrons in $CrI_3$, which in turn could lead to novel topological and correlated states [55-57].

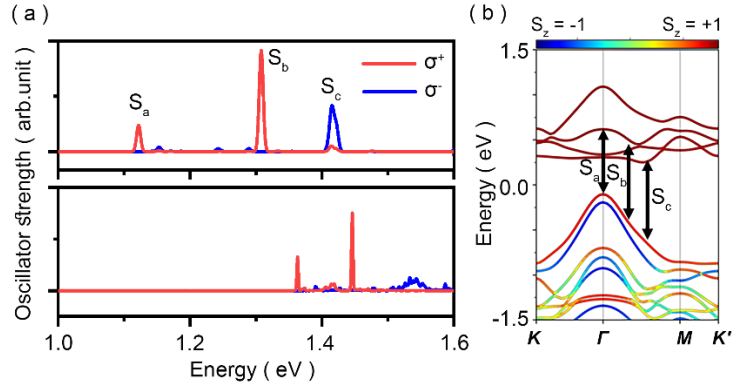

**Fig. 7** (a) Optical absorption for $\sigma^+$ (red) and $\sigma^-$ (blue) circularly polarized photons in a $CrI_3$ monolayer (top) and the twisted $WSe_2/WSe_2/CrI_3$ trilayer (θ = 3.89°, bottom). (b) The spin-polarized band structure of $CrI_3$ monolayer, showing the three intralayer excitons. The Fermi level is set at zero energy.

Lastly, we consider how intralayer excitons in $CrI_3$ may be affected by the imprinted moiré potential. To this end, we compare the optical absorption of intralayer excitons in $CrI_3$ monolayer and the twisted $WSe_2/WSe_2/CrI_3$ trilayer (θ = 3.89°). As shown in **Fig. 7**, three intralayer excitons ($S_a$, $S_b$, and $S_c$) are identified with $S_c$ as excitations from VBM (valence-band-maximum) to CBM (conduction-band-minimum), and $S_a$ and $S_b$ from VBM to CBM+2. We find that $S_a$ and $S_b$ are coupled to $\sigma^+$ photons while $S_c$ to $\sigma^-$ photons, consistent with previous GW-BSE calculations [6,58] for $CrI_3$ monolayer. In the twisted $WSe_2/WSe_2/CrI_3$ trilayer, all three excitons blueshift significantly owing to hole tunneling to the adjacent $WSe_2$ layer (**Fig. S3**). The interlayer electron-hole separation reduces the exciton binding energy (less negative) and increases the exciton energy. Compared to $S_a$ and $S_b$, $S_c$ exhibits larger broadening with reduced oscillator strength because its electron is confined in the MX region by the imprinted moiré potential as shown in **Fig. S4**, thus has little overlap with the hole (which extends to the adjacent $WSe_2$ layer), yielding a smaller oscillator strength and a higher exciton energy. In contrast, the electrons in $S_a$ and $S_b$ have a higher energy (CBM+2) than $S_c$ (CBM), hence are less confined by the imprinted moiré potential.

## III. CONCLUSIONS

We have performed comprehensive first-principles calculations to study the magnetic and moiré proximity effects in $WSe_2/WSe_2/CrI_3$ trilayers. Single-particle calculations are first performed to examine the stacking dependence of valley splitting in non-twisted trilayers. We find that the dominant contribution to the valley splitting is interlayer hybridization that depends on the stacking and the internal electric field. The electric-field dependence of valley splitting shares similar physics to the electrically tunable, valley-selective Feshbach resonances observed in $MoSe_2$ bilayers. TDDFT calculations are performed to determine the optical absorption spectra of the trilayers and from which we examine how the interlayer hybridization, internal electric field and excitonic effect contribute to the valley splitting. In particular, the large valley splitting in MX-$CrI_3$ trilayer is attributed to exchange-induced interlayer tunneling, akin to "super-exchange" coupling, that alters the exciton charge density and binding energy. The moiré Zeeman effect in the twisted trilayers is explored in detail. It is found that valley splitting is magnified in the moiré superlattices thanks to the superposition of Umklapp excitons, giving rise to greater charge redistribution in the moiré excitons. In contrast, interlayer tunneling of the A exciton yields less charge redistribution and a smaller change in the exciton energy of MX-$CrI_3$. The moiré proximity effect is studied by calculating the imprinted moiré potential on target planes and examining its effects on the interlayer and intralayer excitons in $CrI_3$. The imprinted moiré potential effectively modifies the original moiré potential and displaces the moiré excitons to a new set of positions forming a Kagome lattice. Thus, the envisioned feedback from the target layer to the moiré layers is observed, and the cooperation between the direct and imprinted moiré potentials is expected to yield novel topological and correlated physics.

## APPENDIX
**Computational Methods**
**First-principles ground state calculations**

First-principles ground state calculations were performed based on the density functional theory (DFT) with projector-augmented-wave pseudopotentials [59,60] as implemented in Vienna

Ab initio Simulation Package [61,62]. The LDA+U method [63] was used to determine the atomic structures and magnetic properties of CrI$_3$, and the U term is set as 3 eV for Cr atoms. The energy cutoff for the planewave basis set is 300 eV. The atomic structures are fully relaxed until the residual force on each atom is less than 0.01 eV Å$^{-1}$. The energy convergence criterion is set as 1×10$^{-5}$ eV. A vacuum layer of 15 Å is included in the calculations to eliminate spurious interactions between the periodic images of the unit cells. Spin-orbit coupling (SOC) is taken into consideration in the band structure calculations.

**First-principles excited state calculations**

The conventional first-principles approach for determining excitonic properties in semiconductors is GW-Bethe-Salpeter equation (GW-BSE) method [64-66] based on the many-body perturbation theory. However, the GW-BSE method is highly expensive for moiré excitons due to large numbers of atoms in moiré supercells. Inclusion of SOC further increases the computational cost. In this work, we used a recently developed first-principles method that can offer a reliable description of excitons while significantly reducing the computational cost. This approach is based on the linear-response time-dependent density functional theory (TDDFT) [67,68] with optimally tuned (OT), screened and range-separated hybrid (SRSH) exchange-correlation (XC) functionals [34-37]. Unlike the traditional TDDFT methods with local and semi-local XC functionals, the TDDFT-OT-SRSH method can capture the long-range electron-electron and electron-hole interactions in solids correctly by choosing appropriate parameters, and has been extensively used to study optical and excitonic properties in solids and vdW bilayers [38-42,69-74]. Formulated with spinor wavefunctions [33], the TDDFT-OT-SRSH method can also capture noncollinear magnetism, including SOC, in magnetic materials. In this method, the following non-Hermitian eigenvalue equation[75] is solved to determine the exciton energies and wavefunctions:

$$\begin{pmatrix} A & B \\ B^* & A^* \end{pmatrix} \begin{pmatrix} X_I \\ Y_I \end{pmatrix} = \omega_I \begin{pmatrix} 1 & 0 \\ 0 & -1 \end{pmatrix} \begin{pmatrix} X_I \\ Y_I \end{pmatrix} \quad (1)$$

where the pseudo-eigenvalue $\omega_I$ represents the *I*-th exciton energy. The matrix elements of **A** and **B** in the basis of two-component spinor orbitals are given by:

$$A_{ij,kl} = \delta_{i,k}\delta_{j,l}(\varepsilon_j - \varepsilon_i) + K_{ij,kl} \quad (2)$$

$$B_{ij,kl} = K_{ij,lk} \quad (3)$$

Here $K$ is the coupling matrix where indices $i$ and $k$ indicate the occupied Kohn-Sham (KS) orbitals, and $j$ and $l$ represent the virtual KS orbitals. According to the assignment ansatz of Casida, the many-body wavefunction of an excited state $I$ can be written as

$$\Phi_I \approx \sum_{ij} \frac{X_{I,ij}+Y_{I,ij}}{\sqrt{\omega_I}} \hat{a}_j^\dagger \hat{a}_i \Phi_0 = \sum_{ij} Z_{I,ij} \hat{a}_j^\dagger \hat{a}_i \Phi_0 \quad (4)$$

Where $\hat{a}_i$ the annihilation operator acting on the $i$-th KS orbital, and $\Phi_0$ is the ground-state many-body wavefunction taken to be the single-Slater determinant of the occupied KS orbitals. In order to reduce the computational cost associated with the exact exchange on large systems, the first-order perturbation theory to the range-separated hybrid Kohn-Sham Hamiltonian is used [76,77].

In this noncollinear (TD)DFT-OT-SRSH method, there are three parameters α, β and γ that need to be specified. α determines the contribution from the exact exchange and β controls the contribution from the long-range exchange terms. γ is the range-separation parameter. α and β satisfy the requirement of α+β=1/$\varepsilon_0$ where $\varepsilon_0$ is the scalar dielectric constant of the solid. The optimal set of the parameters (α=0.087, β=0.413 and γ=0.12) was determined by fitting the quasiparticle bandgap ($E_g$) of the pristine $WSe_2$ and $CrI_3$ monolayers computed from this method to that of experimental results and GW calculations. With these parameters, the calculated quasiparticle bandgap ($E_g$) and optical gap ($E_{opt}$) are $E_g$=2.236 eV and $E_{opt}$=1.696 eV for pristine $WSe_2$, and $E_g$=1.849 eV and $E_{opt}$=1.02 eV for $CrI_3$, respectively. These results are in good agreement with experimental values ($WSe_2$: $E_g$ = 2.12-2.21 eV, $E_{opt}$ = 1.65-1.73 eV) [78-81] and GW-BSE calculations ($WSe_2$: $E_g$ = 2.10 eV, $E_{opt}$ = 1.58 eV; $CrI_3$: $E_g$ = 2.0 eV, $E_{opt}$ = 1.1 eV) [6].

**Oscillator strength of excitons**

To obtain the optical dipole moment for an exciton, we first determine the single-particle electron-hole transitions involved in the exciton. As mentioned above, the many-body wavefunction of the exciton $\Phi_I$ is expressed as a linear combination of the single-particle electron-hole transitions, and $Z_{I,ij}$ represents the corresponding electron-hole transition amplitude. The polarization-dependent optical dipole moment ($\mu_I$) of the exciton $I$ is calculated as

$$\mu_I = \sum_{ij} Z_{I,ij} P_{ij} \quad (5)$$

$$P_{ij} = \langle \phi_i | \hat{\varepsilon} \boldsymbol{r} | \phi_j \rangle \quad (6)$$

where $P_{ij}$ is the transition dipole moment between the KS occupied state $\phi_i$ and unoccupied state $\phi_j$. $\hat{\boldsymbol{\varepsilon}}$ is the unit vector of the electric field of the polarized light, and $\boldsymbol{r}$ is the position operator of the electron. Since the position operator $\boldsymbol{r}$ is ill-behaved in an extended solid, we use the momentum operator [82,83] to compute the transition dipole moment via the relation $i[\boldsymbol{H},\boldsymbol{r}] = \boldsymbol{p} + i[V_{\text{NL}},\boldsymbol{r}]$, where $\boldsymbol{H}$ is the KS Hamiltonian, $\boldsymbol{p}$ is the momentum operator and $V_{\text{NL}}$ is the nonlocal pseudopotential. Considering that the errors of neglecting the commutator can be significantly reduced by including $d$-projectors in the PAW potential [84,85], we omit the commutator and only consider the momentum operator in the calculation of the transition dipole moment. The oscillator strength is then determined by

$$f = \frac{2m_e \omega_I}{3\hbar^2}\mu_I^2 \quad (7)$$

where $\omega_I$ is the exciton energy, $m_e$ is the mass of the electron and $\hbar$ is the reduced Planck constant.

## ACKNOWLEDGMENTS

This work was supported by the US National Science Foundation (DMR-1828019 and DMR-2105918) and the US Army Research Office (W911NF-23-10205 and W911NF-25-10117).